\documentclass[9pt,reprint,amsmath,amssymb,pra]{revtex4-1}

\usepackage{graphicx}
\usepackage{dcolumn}
\usepackage{bm}

\usepackage{amsfonts}

\begin{document}
%

%
%

\title{Doppler shift generated by a moving diffraction grating under incidence by polychromatic diffuse light}

\author{Kokou B. Dossou}

\altaffiliation{Centre for Ultrahigh-Bandwidth Devices for Optical Systems (CUDOS)
and School of Mathematical and Physical Sciences,
University of Technology Sydney,
PO Box 123,  Broadway,
New South Wales 2007, Australia}

\email{Corresponding author: Kokou.Dossou@uts.edu.au}


 \begin{abstract}
We consider the spectral response of moving diffraction gratings, in which the incident light extends over
a broad angular range and where the diffracted light is observed from a specific angle.
We show that the dispersion relation between the frequency perceived by an observer 
who is looking at a moving grating and the incident frequency can exhibit some unique features, such as a flat band (i.e., a local minimum).
An observer can see the light diffracted into a non-specular diffraction order 
from a multitude of incident light rays and the angle of incidence of each ray is frequency-dependent;
as a consequence, when the grating is moving, each incident ray experiences a Doppler shift in frequency that depends on its angle of incidence.
We find that remarkable features appear near a Wood anomaly where the angle of incidence, for a given 
diffraction angle, can change very quickly with frequency. 
This means that light of multiple frequencies and incident from multiple angles can be mixed 
by the motion of the grating into the same diffracted ray and their frequencies can be compressed
into a narrower range.
The existence of a flat  band means that a moving grating can be used as a device to 
increase the intensity of the perceived diffracted light due to spectral compression.
The properties of a grating in motion in sunlight can also be relevant to the study of naturally
occurring gratings which are typically in oscillatory motion.
\end{abstract}
%

%
%

\maketitle

%
%

\section{Introduction}

The Doppler shift has been successfully used to understand the frequency shift generated 
by moving reflectors, such as a mirror \cite{Sommerfeld:Optics},
a moving grating~\cite{Bahrmann:OC:1977}
or a grating acousto-optic modulator \cite{Korpel:book:1996}.
In many applications, diffraction gratings are typically operated at a fixed incidence angle (and a fixed non-specular diffraction order).
For instance grating spectrometers are used to split a polychromatic ray (at a given incidence angle) into several
diffracted monochromatic rays which propagate in different directions;
the angle of the propagation direction depends on the frequency and can change very quickly with the frequency, near a Wood anomaly.

\begin{figure}[htbp]

 \centerline{
\includegraphics[width=7cm]{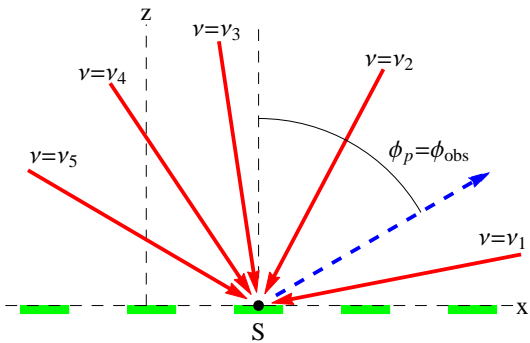}
}
\caption{
Observation at a fixed angle $\phi_{\rm obs}$:
For a given diffraction order $p$, an observer looking at a point $S$ on the grating interface
can see a diffracted light (blue dashed arrow) if its angle of diffraction $\phi_p$ is equal to $\phi_{\rm obs}$.
The angle of the corresponding incident light (red solid arrows) depends on the frequency $\nu$ if $p \neq 0$
(see Eq.~(\ref{phi0:vs:phi:obs})).
}
\label{Fig:observation:angle:linear}
\end{figure}

\medskip
Here we are interested in the problem of an observer who is watching a diffracted ray from a moving grating  (at a fixed non-specular diffraction order).
Since at any given time the observer's viewing angle has a fixed value,
this problem can be seen as a reciprocal situation
of the case where a single polychromatic ray is incident on a grating:
here the observer can only see the diffracted fields which are propagating in a given direction;
the corresponding incident rays are coming from several different directions 
and are merged into a single diffracted ray (see Fig.~\ref{Fig:observation:angle:linear}).
The incidence angle of each incident ray is frequency-dependent and, again,
can change rapidly with frequencies near a Wood anomaly.
In practice, polychromatic continuous diffuse light (e.g., direct sunlight, diffuse skylight) can readily provide 
a light ray with the required incidence angle.

\medskip
The Doppler shift depends on the angle of incidence
and the fact that the angle of incidence is frequency-dependent
implies that, when the grating is moving,  there is variability in the frequency shift 
experienced by each incident frequency.
The main purpose of the current work is to investigate the properties of the Doppler shift created by a moving grating under 
a frequency-dependent incidence angle.
The angular dispersion of the incidence angle with respect to the frequency can be very high 
near a Wood anomaly. As a consequence, the Doppler dispersion relation
(i.e., the observed frequency as a function of the incident frequency)
can be expected to exhibit some interesting non-linear behavior.
In particular, we will look for conditions which lead to the existence of a flat band.
When a band of incident frequencies inside a flat band is Doppler-shifted into a narrower 
 band of observed frequencies (spectral compression), from energy conservation, we can expect that
the observer will perceive an increased intensity of the light diffracted  by a moving
grating compared to the light diffracted by a stationary grating.

\medskip
The properties of a grating in motion in polychromatic diffuse light can also be relevant to the study
of  natural diffraction gratings as they seem to be placed on surfaces designed to undergo oscillatory motion, for instance, 
the wings of butterflies, leaves and flowers of plants
(wind-induced oscillations). 
Many species also display rapid cycles of movement of iridescent colors 
in ritualized dances during courtship,
e.g., bird-of-paradise~\cite{Wilts:PNAS:2014}
and peacock spider~\cite{Otto:Peckhamia:2014}.
Since the work of Robert Hooke~\cite{Hooke:Micrographia:1665}
and Isaac Newton~\cite{Newton:book:Opticks},
it is known that the  iridescent colors from some animals and plant species
are generated by grating structures, although the  functional purpose of the oscillations 
of natural gratings is still not well understood~\cite{Vukusic:CB:2011}.

\medskip
For the numerical simulation of moving gratings, we will consider a grating geometry 
based on the periodic nanostructure found in the wing scales of a \emph{Morpho rhetenor} butterfly.
Our choice is motived by several reasons.
First the blue nanostructure of the \emph{Morpho rhetenor} are one of the most studied butterfly nanostructures.
This nanostructure can also be modelled as a one-dimensional diffraction grating, leading to a substantial reduction
of the computer simulation time and memory usage.
The \emph{Morpho rhetenor} is also well-known for its ability to produce a bright blue color;
for instance, in 1864 the English naturalist and explorer Henry Bates wrote 
that a \emph{Morpho rhetenor} butterfly flapping its wings in the sunlight can produce 
a blue flash that ``is visible a quarter of a mile off''~\cite{Bates:book:1864}
and it has been reported that this flash can even be seen from a low-flying aircraft~\cite[see p. 218]{Silberglied:SRSL:1984}.
The relatively long range of visibility of the \emph{Morpho} butterfly wing flap is consistent with our prediction 
that the motion of a grating under sunlight can induce a strong reflection due to the spectral compression
and so the grating model for the \emph{Morpho rhetenor} nanostructure is an appropriate example for a case study.

\medskip
The paper is organized as follows.
In Section~\ref{section:Angular:dispersion}, we will analyze the angular dispersion of the incidence angle
with respect to frequency.
The properties of the Doppler shift produced by a grating in a translational motion
and a rotational motion will be investigated, respectively, in
Sections~\ref{section:translational:motion} and \ref{section:rotational:motion}.
Some numerical simulation results will be presented in Section~\ref{section:numerical:simulations}
and a conclusion section will follow.

%
%

%
%

%
%

\section{Angular dispersion at a fixed angle of diffraction}

\label{section:Angular:dispersion}

\begin{figure}[htbp]
 \centerline{
\includegraphics[width=7cm]{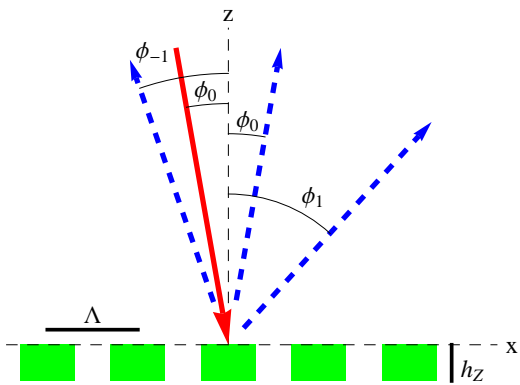}
}
\caption{Illustration of a diffraction grating.
The grating is periodic in the $x$-direction, with a period $\Lambda$, 
invariant in the $y$-direction and has a finite thickness $h_z$ in the $z$-direction.}

\label{Fig:grating}
\end{figure}

\begin{figure}[htbp]

 \centerline{
\includegraphics[width=5cm]{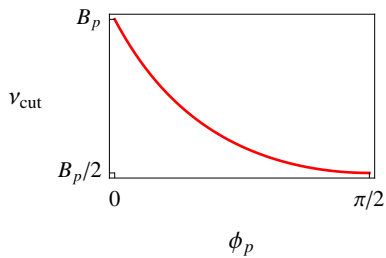}
}
\caption{
Curve of the cut-off frequency  $\nu_{\rm cut}$ in Eq.~(\ref{cut-off:p:positive}) as a function 
of the angle of diffraction $\phi_p$ (for $p>0$).
}
\label{Fig:cutoff:order:p}
\end{figure}

As a model problem, we consider a one-dimensional diffraction grating
as illustrated in Fig.~\ref{Fig:grating}.
The grating is periodic in the $x$-direction, with a period $\Lambda$,
invariant with respect to $y$ and has a finite thickness in the $z$-direction.
The analysis of the angular dependence is based on the grating equation 
\begin{eqnarray}
\label{diffrac:eq1}
\sin \phi_p & = & \sin \phi_0 + \frac{p \, \lambda}{\Lambda} ,
\end{eqnarray}
where $\lambda$ is the wavelength,
$p$ is a given diffraction order,
$\phi_0$ and $\phi_p$ are respectively the angle of incidence and the angle of diffraction.
We consider an incidence by a continuous polychromatic diffuse light
with an observer who is looking at the grating at an observation angle $\phi_{\rm obs}$.
For a given incident field with a wavelength $\lambda$, the observer
can see the light diffracted into the order $p$ when $\phi_p = \phi_{\rm obs}$
and the corresponding angle of incidence $\phi_0$ can be obtained from the grating equation Eq.~(\ref{diffrac:eq1}):
\begin{eqnarray}
\label{phi0:vs:phi:obs}
 \phi_0 
& = & {\rm arcsin} \left( \sin \phi_p - \frac{B_p}{\nu} \right) ,
\end{eqnarray}
where $\nu = c/\lambda$ is the frequency of the incident light,
$c$ being the speed of light in vacuum.
The symbol $B_p$ is defined as
\begin{eqnarray}
\label{param:b}
B_p & = & \frac{c \, p}{\Lambda}.
\end{eqnarray}
The angular dispersion of the angle of incidence with respect to the frequency $\nu$
can be derived from Eq.~(\ref{phi0:vs:phi:obs}):
\begin{eqnarray}
\label{angular:dispersion}
\frac{d \phi_0}{d \nu} 
= \frac{B_p}{\nu^2  \, \sqrt{1 - \left( \sin \phi_p - \frac{B_p}{\nu} \right)^2}}
= \frac{B_p}{\nu^2  \, \cos \phi_0},
\end{eqnarray}
since $\sin \phi_p - B_p/ \nu = \sin \phi_p - p \, \lambda / \Lambda = \sin \phi_0$
(see Eq.~(\ref{diffrac:eq1})).
The incidence angle Eq.~(\ref{phi0:vs:phi:obs})
exists as a real number $\phi_0 \in \mathbb{R}$ if and only if
$-1 \leq \sin \phi_p - p \,\lambda / \Lambda \leq 1$,
i.e., if $ p > 0$, we have
\begin{eqnarray}
\label{cut-off:p:positive}
\lambda \leq \lambda_{\rm cut}
= (1 + \sin \phi_p) \, \frac{\Lambda}{p} 
 \Longleftrightarrow 
\nu \geq \nu_{\rm cut} =  \frac{B_p}{1 + \sin \phi_p} ,
\end{eqnarray}
while if $p < 0$, we get
\begin{eqnarray}
\label{cut-off:p:negative}
\lambda \leq \lambda_{\rm cut}
= (1 - \sin \phi_p) \, \frac{\Lambda}{-p}
 \Longleftrightarrow 
\nu \geq \nu_{\rm cut} 
= \frac{-B_p}{1 - \sin \phi_p} .
\end{eqnarray}
The curve in Fig.~\ref{Fig:cutoff:order:p} shows  the profile of the cut-off frequency $\nu_{\rm cut}$ as the  angle of diffraction $\phi_p$ varies
(for the case $p>0$). 
The angle of incidence at the cut-off frequency is either $\phi_0 = - \pi/2$ if $p>0$ or 
$\phi_0 = \pi/2$ if $p<0$.
This shows that the angular dispersion Eq.~(\ref{angular:dispersion}) can be arbitrarily large when 
$\nu$ is near the cut-off frequency $\nu_{\rm cut}$ defined in Eq.~(\ref{cut-off:p:positive}) or Eq.~(\ref{cut-off:p:negative}). 
The following asymptotic relation is valid for a frequency $\nu$ near $\nu_{\rm cut}$:
\begin{eqnarray}
\label{incidence:diffuse:approx}
\phi_0
& \approx &
- {\rm sign}(p) \left(
\frac{\pi}{2} - \frac{\sqrt{2 \, |B_p| \, (\nu - \nu_{\rm cut})}}{\nu_{\rm cut}}  
\right) ,
\end{eqnarray}
where ${\rm sign}(p) = 1$ if $p>0$ and ${\rm sign}(p) = -1$ if $p<0$.
The square root function in Eq.~(\ref{incidence:diffuse:approx}) shows that the incidence angle $\phi_0$
can vary rapidly with frequencies $\nu$ near $\nu_{\rm cut}$.
The relation Eq.~(\ref{incidence:diffuse:approx}) can be derived from
Eq.~(\ref{phi0:vs:phi:obs})
by using the fact that $B_p = {\rm sign}(p) \, |B_p|$
and $\sin \phi_p =  B_p/\nu_{\rm cut} - {\rm sign}(p) $
(since $\phi_0 = - {\rm sign}(p) \, \pi/2$ at the cut-off frequency),
and by applying the asymptotic relation 
${\rm arcsin} \, \left( 1 - x \right)  \approx \frac{\pi}{2} - \sqrt{2 \, x}$
for small values of $x$.
%

%
%

%
%

%
%

%
%

%
%

\section{Frequency shifting by a grating in a translational motion}

\label{section:translational:motion}

We assume that the grating is in a translational motion with a velocity parallel to the normal direction $\bm{z}$.
As illustrated in Fig.~\ref{Fig:observation:angle:linear},
an observer, who is looking at a point $S$ on the grating at a viewing angle $\phi_{\rm obs}$,
can see the light diffracted into an order $p$ if the angle of diffraction is equal to the observation angle:
\begin{eqnarray}
\label{eq1:observation:angle}
\phi_p = \phi_{\rm obs} ,
\end{eqnarray}
and the corresponding angle of incidence $\phi_0$ is given by Eq.~(\ref{phi0:vs:phi:obs}).
The angle $\phi_0$ is frequency-dependent for non-specular orders.
Note that we will consider in Section~\ref{section:rotational:motion}\ref{subsection:different:phi}, some particular cases
where the condition Eq.~(\ref{eq1:observation:angle}) can be relaxed so that
$\phi_p$ can be different from $\phi_{\rm obs}$.

\medskip
Translational motion of the grating will create a Doppler shift and we want to compute the frequency shift perceived by the observer.
The reflection of a wave from a moving surface involves a double Doppler shift.
First at incidence, the point $S$ can be seen as a receiver moving at the velocity $\bm{V} = V \, \bm{z}$, 
the velocity component in the direction of the light source is given by $V \, \cos \phi_0$
so that the Doppler shift (in the classical limit where $|V/c| \ll 1$) is 
\begin{eqnarray}
\label{Doppler:shift:0:linear}
\Delta \nu_0 = \nu \, \frac{V \, \cos \phi_0}{c} .
\end{eqnarray}
On reflection, the point $S$ can be seen as a light source moving at the speed $V$; 
the velocity component in the observer's direction is given by $V \, \cos \phi_p$
so that the Doppler shift is:
\begin{eqnarray}
\label{Doppler:shift:p:linear}
\Delta \nu_p = \nu \, \frac{V \, \cos \phi_p}{c} .
\end{eqnarray}
The total Doppler shift is
\begin{eqnarray}
\label{total:Doppler:shift:linear}
\Delta \nu = \Delta \nu_0 + \Delta \nu_p
= \nu \, \frac{V \, \left(\cos \phi_0 + \cos \phi_p \right)}{c} ,
\end{eqnarray}
and the observed frequency can be written as
\begin{eqnarray}
\label{dispersion:eq1a:linear}
\nu'
= \nu + \Delta \nu
= \nu + \nu \, A \, \left(\cos \phi_0 + \cos \phi_p \right) ,
\end{eqnarray}
with
\begin{eqnarray}
\label{param:a:linear}
A  & = & \frac{V}{c} .
\end{eqnarray}
For the specular order $p=0$, we have $\phi_0 = \phi_p$ and Eq.~(\ref{total:Doppler:shift:linear}) takes the form
$\Delta \nu = - 2\, \nu \, A \, \cos \phi_0$,
which is the formula of the Doppler shift created by a flat mirror in a translational motion~\cite[p.~74]{Sommerfeld:Optics}.
The Doppler shift Eq.~(\ref{total:Doppler:shift:linear}) is also in agreement with the formula for the frequency shift  by a moving grating
in \cite[see Eq.~(1)]{Bahrmann:OC:1977}; note that the grating considered in \cite{Bahrmann:OC:1977} was moving along
the $x$-direction and so the cosine function in Eq.~(\ref{total:Doppler:shift:linear}) is replaced by the sine function in \cite{Bahrmann:OC:1977}.

\medskip
There are some important differences between the Doppler shift from a  specular order and a non-specular order.
For the case of specular reflection, the angle of incidence $\phi_0$ is equal to $\phi_p$, i.e., independent of the frequency,
and there is no cut-off frequency;
in particular the Doppler shift scales linearly with the frequency.
In a contrast, with a non-specular reflection, the angle of incidence depends on the frequency and the diffracted field can propagate only for frequencies
above a cut-off frequency $\nu_{\rm cut}$;
this means that, for an illumination by a diffuse polychromatic light, the Doppler shift for non-specular orders can display some unique and interesting  behaviors.

\begin{figure}[htbp]

 \centerline{
\includegraphics[width=8.5cm]{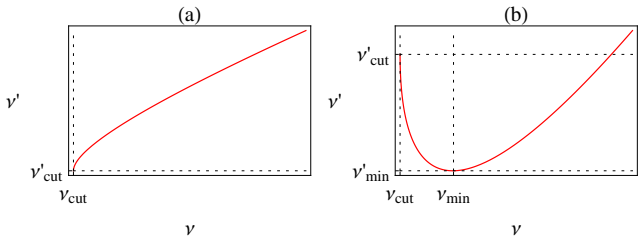}
}
\caption{Typical profiles of the non-specular Doppler dispersion relation Eq.~(\ref{dispersion:eq1a:linear})
near a cut-off.
The velocity $V$ is respectively positive and negative for the panels (a) and (b).
}
\label{Fig:dispersion:a:linear}
\end{figure}

\medskip
For a non-specular order $p$,
the application of the asymptotic relation Eq.~(\ref{incidence:diffuse:approx}) together with
$\cos(\pi/2 - x) = \sin x \approx x$,
leads to the following 
asymptotic expression of Eq.~(\ref{dispersion:eq1a:linear}) 
for frequencies $\nu$ near $\nu_{\rm cut}$:
\begin{eqnarray}
\label{dispersion:approx1:linear}
\nu' 
& \approx &
\nu 
+ \nu \, A \, \cos \phi_p
+ A 
\sqrt{
2 \, |B_p| \, \left( \nu - \nu_{\rm cut} \right)
} ,
\end{eqnarray}
where the estimate $\nu / \nu_{\rm cut} \approx 1$
is used for the factor of the square root function.
The presence of the square root function in Eq.~(\ref{dispersion:approx1:linear})
indicates that the observed frequency $\nu'$ changes rapidly when 
the incident frequency $\nu$ varies near the cut-off $\nu_{\rm cut}$.
In fact near $\nu_{\rm cut}$, the derivative of the shift 
$\Delta \nu = 2 \,  A \, 
\sqrt{
2 \, |B_p| \, \left( \nu - \nu_{\rm cut} \right)
}$
with respect to the frequency $\nu$ dominates that of the linear terms $\nu (1 + V \, \cos \phi_p / c)$
and so the square root term determines the sign of the slope of 
the Doppler dispersion relation $\nu' = \nu + \Delta \nu$ near a cut-off frequency.
If $V > 0$, 
i.e., the grating is moving toward the observer,
the observed frequency  $\nu'$ increases monotonically with 
the frequency $\nu$ (see Fig.~\ref{Fig:dispersion:a:linear}~(a)).
As illustrated in  Fig.~\ref{Fig:dispersion:a:linear}~(b),
if $V < 0$, 
i.e., the grating is moving away from the observer,
 the asymptotic function Eq.~(\ref{dispersion:approx1:linear})
is a decreasing function when the incident frequency $\nu$ is close enough 
to the cut-off $\nu_{\rm cut}$, reaches a minimum and start increasing 
as the contribution of the linear term dominates when $\nu$  is far enough from $\nu_{\rm cut}$.
A detailed analysis of the radiated power for frequencies near the minimum will be carried out
in the next section (for the case of a rotating grating) and will not be repeated here as
the results for a rotational motion and a translational motion are very similar.

\medskip
We note that  if the grating is moving horizontally  instead of vertically
then the observed frequency Eq.~(\ref{dispersion:eq1a:linear}) takes the form
$\nu' = \nu + \nu \, A \, \left(\sin \phi_p - \sin \phi_0 \right)$
and the asymptotic expression Eq.~(\ref{dispersion:approx1:linear}) becomes
$\nu'  \approx \nu 
+ \nu \, A \, \sin \phi_p
\pm A \left( \nu - |B_p| \, \left( \nu - \nu_{\rm cut} \right) \right)$,
since we have 
$\sin \phi_0 = \pm \sqrt{1 - \cos^2 \phi_0}
\approx\pm( 1 - (\cos^2 \phi_0)/2)$.
We can conclude that the observed frequency does not have a flat band
since the square root term does not appear in the asymptotic expression.
%


%
%

%
%

%
%

\section{Frequency shifting by a grating in a rotational motion}

\label{section:rotational:motion}

%
%
%

We now suppose that the grating depicted in Fig.~\ref{Fig:observation:angle:linear}
is rotating around the $y$-axis.
For a given incident ray, as the grating is rotating,
the angle of incidence $\phi_0$ of the ray
will change with an angular velocity denoted $\Omega_0$.
For a given diffraction order $p$, the diffracted wave will also be rotating
at an angular speed $\Omega_p = d \phi_p/ dt$, the value of which can be obtained by differentiating 
Eq.~(\ref{diffrac:eq1}).
The time-derivative of the grating equation  Eq.~(\ref{diffrac:eq1}),
at fixed frequency, leads to the following angular momentum conservation relation:
\begin{eqnarray}
\label{angular:momentum}
\Omega_p \, \cos \phi_p & = & \Omega_0 \, \cos \phi_0 ,
\end{eqnarray}
so the angular velocity $\Omega_p$ can be expressed in term of $\Omega_0$ as
\begin{eqnarray}
\label{angular:velocity}
\Omega_p & = &
\frac{\cos \phi_0}{\cos \phi_p} \, \Omega_0 .
\end{eqnarray}
We now apply a treatment similar to the one presented in Section~\ref{section:translational:motion},
in order to find the Doppler-shifted frequency perceived by an observer watching
the point $S=(x_S,0)$ on the grating interface $z = 0$.
At incidence, the point $S$ can be seen as a receiver moving at the speed
\begin{eqnarray}
\label{speed:V:0}
\bm{V}_0 = - x_S \, \Omega_0 \, \bm{z} , 
\end{eqnarray}
the velocity component in the direction of the light source is given by $V_0 \, \cos \phi_0$.
The minus sign in the expression$\bm{V}_0 = - x_S \, \Omega_0 \, \bm{z}$ comes from the
fact that when $\Omega_0>0$, a point $S$ on the left of the rotation axis moves upward while it moves downward on the right side.
At reflection, since the diffracted field is rotating at the  angular speed $\Omega_p$, 
the point $S$ can be seen as a light source moving at the speed
\begin{eqnarray}
\label{speed:V:p}
\bm{V}_p = - x_S \, \Omega_p \, \bm{z} ,
\end{eqnarray}
the component of $\bm{V}_p$ in the observer's direction is $V_p \, \cos \phi_p$.
The total Doppler shift is then
\begin{eqnarray}
\label{total:Doppler:shift}
\Delta \nu = \Delta \nu_0 + \Delta \nu_p
= - \nu \, \frac{x_S \, \left( \Omega_0 \, \cos \phi_0 + \Omega_p \, \cos \phi_p \right)}{c} .
\end{eqnarray}
For the specular order $p=0$, Eq.~(\ref{total:Doppler:shift}) takes the form
$\Delta \nu = - 2\, \nu \, x_S \, \Omega_0 \, \cos \phi_0 / c$,
which is the formula of the Doppler shift created by a rotating flat mirror.
%
By applying the angular momentum relation Eq.~(\ref{angular:momentum}), we can verify that 
Eq.~(\ref{total:Doppler:shift}) also takes the form $\Delta \nu = - 2\, \nu \, x_S \, \Omega_0 \, \cos \phi_0 / c$,
for non-specular orders.
However, for a non-specular order, the angle of incidence $\phi_0$ is frequency-dependent and a cut-off exists;
again, this means that for a rotating grating illuminated by a diffuse polychromatic light, the wave diffracted into a non-specular order
can display some non-trivial Doppler effect which will be studied at the next section.
%

%
%

\subsection{Asymptotics of the frequency shift}

An asymptotic treatment can provide some useful insight into the physics of a moving grating in a polychromatic diffuse light.
From Eq.~(\ref{total:Doppler:shift}) and the angular momentum relation Eq.~(\ref{angular:momentum}),
the observed frequency can be written as
\begin{eqnarray}
\label{dispersion:eq1}
\nu' = \nu + \Delta \nu
 =  \nu + 2 \, \nu \, A \, \cos \phi_0 ,
\end{eqnarray}
with
\begin{eqnarray}
\label{param:a}
A  & = & \frac{- x_S \, \Omega_0}{c} .
\end{eqnarray}
By repeating the derivation for the asymptotic approximation Eq.~(\ref{dispersion:approx1:linear}),
we can obtain the following asymptotic expression of 
the observed frequency Eq.~(\ref{dispersion:eq1})
\begin{eqnarray}
\label{dispersion:approx1}
\nu' 
& \approx &
\nu + 2 \,  A \, 
\sqrt{
2 \, |B_p| \, \left( \nu - \nu_{\rm cut} \right)
} .
\end{eqnarray}

\medskip
Numerical tests have shown that the asymptotic approximation Eq.~(\ref{dispersion:approx1}) accurately reflects the behavior 
of the Doppler dispersion relation  Eq.~(\ref{dispersion:eq1}) near a cut-off.
Accordingly we now carry out an analytic investigation of Eq.~(\ref{dispersion:eq1}) based on the asymptotic Eq.~(\ref{dispersion:approx1}).
If $A > 0$, the observed frequency  $\nu'$ increases monotonically with 
the frequency $\nu$ (see Fig.~\ref{Fig:dispersion:a:linear}~(a)).
As illustrated in  Fig.~\ref{Fig:dispersion:a:linear}~(b),
if $A < 0$,  the asymptotic function Eq.~(\ref{dispersion:approx1})
is a decreasing function when the incident frequency $\nu$ is close enough 
to the cut-off $\nu_{\rm cut}$.
We can verify that the asymptotic function
reaches a minimum when $\nu$ takes the value
\begin{eqnarray}
\label{nu:min}
\nu_{\rm min} & \approx & \nu_{\rm cut} + 2 \, A^2 \, |B_p| ,
\end{eqnarray}
and the corresponding observed frequency is
\begin{eqnarray}
\label{nu:p:min}
\nu'_{\rm min} \approx 
\nu_{\rm min} + 2 \, A \, 
\sqrt{
2 \, |B_p| \left( \nu_{\rm min} - \nu_{\rm cut} \right)
}
\approx \nu_{\rm cut} - 2 \, A^2 |B_p| .
\end{eqnarray}
We note that the cut-off frequency $\nu_{\rm cut}$ is the average of
$\nu'_{\rm min}$ and $\nu_{\rm min}$.
With the parameters considered in this work, 
the frequency width $(\nu_{\rm min} - \nu_{\rm cut}) \approx 2 \, A^2 \, |B_p|$ is very small compared to $\nu_{\rm min}$
or $ \nu_{\rm cut} \in \left[ |B_p|/2, |B_p|  \right]$ (see Eqs.~(\ref{cut-off:p:positive}) and (\ref{cut-off:p:negative})).
As an illustration, if the velocity of the point $S$ is $V = 10 \, \rm{m/s}$, then we have
$(\nu_{\rm min} - \nu_{\rm cut}) / \nu_{\rm cut} \leq 4 \, A^2 = 4 \, (V/c)^2 =4.4 \times 10^{-15}$.
We will show in Section~\ref{section:rotational:motion}\ref{subsection:different:phi} that a larger value for the parameter $A$ 
can be possible (see Eq.~(\ref{param:a:scale}))
and this leads to an increased value of the ratio $(\nu_{\rm min} - \nu_{\rm cut}) / \nu_{\rm cut}$.
The cut-off frequency curve in Fig.~\ref{Fig:cutoff:order:p} can also be relevant to the frequencies $\nu_{\rm min}$ and $\nu'_{\rm min}$ 
when they are close to $\nu_{\rm cut}$.

%
%

\subsection{Diffraction efficiency for frequencies $\nu$ near $\nu_{\rm min}$}

\label{subsection:diffraction:efficiency}

For scattering problems, the diffraction efficiency
\begin{eqnarray}
\label{diffraction:efficiency}
e_p = \frac{|\rho_{p}^2| \, \cos \phi_p}{\cos \phi_0}
\end{eqnarray}
into an order $p$ is a quantity of interest
and it needs to be sufficiently high in order to have a physically detectable diffracted light.
The symbol $\rho_{p}$ in Eq.~(\ref{diffraction:efficiency}) is the diffraction coefficient into the order $p$.
For a stationary grating, the diffraction efficiency is usually defined for each frequency.
However with a rotating grating, the incident frequencies can be Doppler-shifted 
in a non-uniform way (especially inside a flat band), and so it can be useful to characterize the diffraction efficiency
in terms of integrals of the radiated powers over a given frequency band.

\medskip
For frequencies $\nu$ in the vicinity of $\nu_{\rm min}$, the asymptotic function  Eq.~(\ref{dispersion:approx1})
can be replaced by its second order Taylor series (see Eq.~(\ref{nu:min})):
\begin{eqnarray}
\label{dispersion:approx:P2}
\nu' 
 \approx 
\frac{(\nu - \nu_{\rm min})^2}{8 \, A^2 \, |B_p|}
+ \nu'_{\rm min}
 \approx 
\frac{(\nu - \nu_{\rm min})^2}{4 \, (\nu_{\rm min} - \nu_{\rm cut})}
+ \nu'_{\rm min} .
\end{eqnarray}
The parabolic approximation~Eq.~(\ref{dispersion:approx:P2}) shows that 
a frequency band centered on $\nu_{\rm min}$ and of width $\mathcal{W} \ll 2 \, (\nu_{\rm min} - \nu_{\rm cut})$,
i.e., $\nu\in [\nu_{\rm min} - \mathcal{W}/2, \nu_{\rm min} + \mathcal{W}/2]$,
will be Doppler-shifted into the band 
$\nu'  \in [\nu'_{\rm min}, \nu'_{\rm min} + \mathcal{W}']$,
with $\mathcal{W}' = \mathcal{W}^2/(16 \, (\nu_{\rm min} - \nu_{\rm cut}))$.
The ratio of the bandwidths $\mathcal{W}$ and $\mathcal{W}'$  is:
\begin{eqnarray}
\label{compression:ratio}
f_p = \frac{\mathcal{W}}{\rule{0cm}{0.35cm} \mathcal{W}'}
=
\frac{16 \, (\nu_{\rm min} - \nu_{\rm cut})}{\mathcal{W}}
=
\sqrt{\frac{16 \, (\nu_{\rm min} - \nu_{\rm cut})}{\rule{0cm}{0.35cm} \mathcal{W}'}}.
\end{eqnarray}
The energy flow carried through the upper interface of a grating unit cell by an incident plane wave with intensity $I$ is proportional to $(I  \, \cos \phi_0)$
and the energy flow of the diffracted wave of order $p$  is proportional to  $(I  \, |\rho_{p}^2| \, \cos \phi_p)$.
For incidence by plane waves with a frequency $\nu$ in the band $[\nu_{\rm min} - \mathcal{W}/2, \nu_{\rm min} + \mathcal{W}/2]$,
the normalized total energies radiated
by the incident waves and diffracted waves are respectively  
\begin{eqnarray}
\label{total:energy:0}
\mathcal{E}_0 & = & 
\int_{\nu_{\rm min} - \mathcal{W}/2}^{\nu_{\rm min} + \mathcal{W}/2}
I(\nu)  \, \cos \left(\phi_0(\nu)\right) d\nu ,  \\
\label{total:energy:p}
\mathcal{E}_p & = &  
\int_{\nu_{\rm min} - \mathcal{W}/2}^{\nu_{\rm min} + \mathcal{W}/2}
I(\nu) \, |\rho_{p}^2(\nu)| \,  \, \cos \left(\phi_p(\nu)\right) d\nu .
\end{eqnarray}
If we assume that the width $\mathcal{W}$ is small enough so that the integrals
Eqs.~(\ref{total:energy:0}) and (\ref{total:energy:p}) can be evaluated by the 
midpoint numerical integration formula, we then have
\begin{eqnarray}
\mathcal{E}_0 & \approx & \mathcal{W} \, I  \, \cos \phi_0 , \\
\mathcal{E}_p & \approx & \mathcal{W} \, I  \, |\rho_{p}^2| \, \cos \phi_p ,
\end{eqnarray}
where $\phi_0 = \phi_0(\nu_{\rm min})$,
$I = I(\nu_{\rm min})$,
$\rho_{p} = \rho_{p}(\nu_{\rm min})$.

\medskip
For the case of a rotating grating, at a given time $t$,
the assumption can be made that the power $\mathcal{E}_p'$ radiated by a diffracted field
over the Doppler-shifted frequency band $[\nu'_{\rm min}, \nu'_{\rm min} + \mathcal{W}']$
is same as the power $\mathcal{E}_p$ radiated by a stationary grating:
\begin{eqnarray}
\label{total:energy:p:prime}
\mathcal{E}_p' = \mathcal{E}_p \approx \mathcal{W} \, I  \, |\rho_{p}^2| \, \cos \phi_p .
\end{eqnarray}
We could define an effective diffraction efficiency as the ratio of the diffracted power $\mathcal{E}_p'$ 
by the incident power over the frequency band $[\nu'_{\rm min}, \nu'_{\rm min} + \mathcal{W}']$.
But we intend to compare the reflectance of a moving grating against that of a perfectly reflecting mirror
(both have same size),
as any reflector with an effective reflectance equal or higher than that of a perfect mirror can be expected
to produce exceptionally bright reflected light.
Accordingly, we will compare $\mathcal{E}_p'$ to the maximum incident power flow
(i.e., $\cos \phi_0 = 1$) for frequencies $\nu' \in  [\nu'_{\rm min}, \nu'_{\rm min} + \mathcal{W}']$:
\begin{eqnarray}
\mathcal{E}_0' = 
\int_{\nu'_{\rm min}}^{\nu'_{\rm min} + \mathcal{W}'}
I(\nu) \, d\nu
\approx \mathcal{W}' \, I .
\end{eqnarray}
The ratio of $\mathcal{E}_p'$ and $\mathcal{E}_0'$ can be written as (see Eq.~(\ref{compression:ratio})):
\begin{eqnarray}
\label{power:ratio}
\mathcal{D}_p' 
= \frac{\mathcal{E}_p'}{\mathcal{E}_0'}
\approx \frac{\mathcal{W} \, |\rho_{p}^2| \, \cos \phi_p}{\mathcal{W}'}
\approx 
4 \, |\rho_{p}^2| \, \cos \phi_p
\sqrt{\frac{\nu_{\rm min} - \nu_{\rm cut}}{\rule{0cm}{0.35cm} \mathcal{W}'}} .
\end{eqnarray}
%

%
%

\subsection{Visual range of a moving grating}

\label{section:Visual:range}

The visual range of a reflector  (i.e., the maximum distance at which the reflector can be seen)
can also be used to characterize the strength of a reflected radiation.
Indeed the reflectivity of the \emph{Morpho rhetenor} butterfly 
is often described in term of a distance where it can be seen~\cite{Bates:book:1864,Silberglied:SRSL:1984}.
Here we assume that a perfect mirror is observed at a  distance equal to its visual range,
and by using a simplified but realistic model we will identify conditions under which the diffracted light from a moving
can also be seen at the same distance.
We note that, during daytime, the background radiation from the sky or the sun is the main limiting factor for the visual range.

\medskip
The power ratio Eq.~(\ref{power:ratio}) takes the value $\mathcal{D}_p' = 4$, for instance, when $\mathcal{W}'$ is set to
\begin{eqnarray}
\label{super:efficiency}
\mathcal{W}'
& = &
(\nu_{\rm min} - \nu_{\rm cut}) \,
|\rho_{p}^4| \, \cos^2 \phi_p ,
\end{eqnarray}
in theory, this means that the moving grating can reflect a brighter light than a perfect passive reflector with $100\%$ reflectance, 
for observed frequencies $\nu' \in  [\nu'_{\rm min}, \nu'_{\rm min} + \mathcal{W}']$.
If a moving diffraction grating can radiate at least 4 times more power than a perfect reflector
over the observed frequency range $\nu' \in  [\nu'_{\rm min}, \nu'_{\rm min} + \mathcal{W}']$,
then it can be physically possible to see such a relatively bright light at a distance greater than 
the visual range of a perfect reflector.
The brightest light will occur 
for observed frequencies $\nu' \in  [\nu'_{\rm min}, \nu'_{\rm min} + \mathcal{W}']$
where the incident light comes directly from the sun; since the grating is rotating the light will then appear 
as a sudden bright flash.
%

%
%

\begin{figure}[htbp]

 \centerline{
\includegraphics[width=8cm]{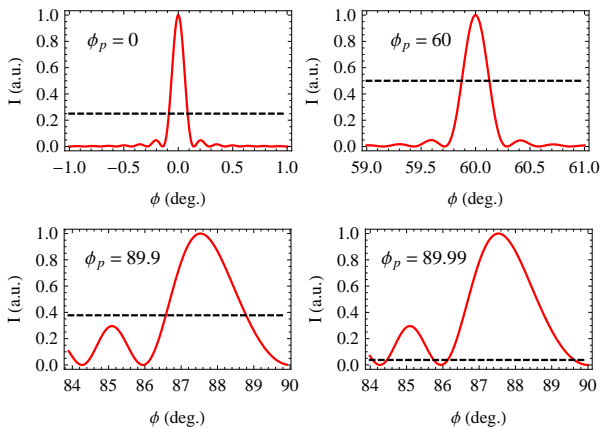}
}
\caption{Diffracted intensity $I (\phi)$ obtained  with the Fraunhofer diffraction formula Eq.~(\ref{Fraunhofer:formula})
for the angles of diffraction $\phi_p \in \{0^{\circ}, 60^{\circ}, 89.9^{\circ}, 89.99^{\circ} \}$.
The dashed curve is the intensity $I(0)$ under a normal incidence
($I(0)$ is the maximum intensity at a given distance $R$).
}
\label{Fig:Fraunhofer:diffraction}
\end{figure}

%
%

\medskip
In order to gain some qualitative insight, let us assume that the reflected fields are monochromatic plane waves emerging 
from a finite grating of length $L$.
If the grating is conceptualized as an aperture of width $L$ (the interface is located at $z=0$ with $x \in [-L/2,L/2]$), the field
at a point $(x,y) = (R \sin \phi, R \cos \phi)$ can be calculated using 
the Fraunhofer diffraction formula (far-field approximation)~\cite[see Eq.~(5)]{Ferrer:NIMPRb:1995}, \cite{Hecht:book:1998}:
\begin{eqnarray}
\label{Fraunhofer:formula}
I (\phi)
 = 
I_0 \, \frac{L^2 \, \cos^2 \phi}{\lambda \, R} \, 
\left(
{\rm sinc} \left( 
\left( \sin \phi - \sin \phi_p \right)
\frac{\pi \, L}{\lambda} \right)
\right)^2 ,
\end{eqnarray}
where $I_0$ is the intensity of the plane wave at the aperture interface.
The intensity curves are plotted in Fig.~\ref{Fig:Fraunhofer:diffraction} (red continuous curves)
for the angles of diffraction $\phi_p = 0^{\circ}$, $\phi_p = 60^{\circ}$, $\phi_p =89.9^{\circ}$
and $\phi_p =89.99^{\circ}$.
%
For the numerical simulation in Fig.~\ref{Fig:Fraunhofer:diffraction},
the aperture of length is set to $L = 200 \, \mu{\rm m}$ 
(as a one dimensional model of the scales of \emph{Morpho rhetenor} butterflies which are typically $75 \times 200 \mu{\rm m}$ in size~\cite{Vukusic:PCSb:1999})
and the incident wavelength is set to $\lambda = 500 \, \rm{nm}$ (blue light).
The horizontal dashed lines in Fig.~\ref{Fig:Fraunhofer:diffraction} represent the intensity level
$I(0) = I_0 \, L^2 / (\lambda \, R)$
which is the maximum value of $I (\phi)$  for given values of $I_0$ and $R$
(the maximum then occurs at a normal angle of propagation $\phi_p = 0$ for $\phi = 0$).
For each of the panels  in Fig.~\ref{Fig:Fraunhofer:diffraction}, the power flow at the grating interface
from the field associated with the red continuous curve is four times that of the field associated with the dashed line 
(or the intensity $I_0$ in Eq.~(\ref{Fraunhofer:formula}) is set, respectively for the continuous and dashed curves, 
to $I_0 = I_A$ and $I_0 = I_B$ such that $I_A = 4 \, I_B / \cos \phi_p$).
So if $R$ is the visual range of a perfect reflector, the portion of the continuous curve which is above the dashed curve,
corresponds to the directions where the field radiated by the grating can be seen beyond the visual range of a perfect reflector.
We note that for the high grazing angle of diffraction  $\phi_p =89.99^{\circ}$, the intensity $I_A = 4 \, I_B / \cos \phi_p$ 
is very large compared to $I_B$ and if the power flow associated with the red continuous curve is six time smaller (instead of four times larger)
than the power flow associated with the dashed line,
i.e., $I_A = I_B / (6 \, \cos \phi_p)$, the peak intensity of the red continuous curve will still be above the dashed line.
So for grazing angles of diffraction, a diffracted light can still be visible beyond the visual range of a perfect reflector
even if the power ratio $\mathcal{D}_p'$ is a small fraction of one.
%

%
%

\begin{figure}[htbp]

 \centerline{
\includegraphics[width=6cm]{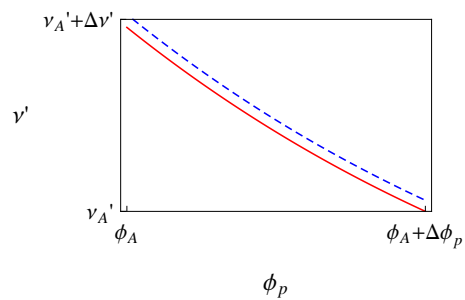}
}
\caption{Illustration of the range of angles and frequencies where a detector collects
the diffracted light.
The red continuous curve and the blue dashed curve represent respectively
the frequencies $\nu'_{\rm min}$ and ($\nu'_{\rm min} + \mathcal{W}')$
as a function of $\phi_p$.}
\label{Fig:detector:range}
\end{figure}

%
%

\medskip
In practice, when a photodetector (e.g., eye)  is used to detect the reflected light from the grating,
it can collect the diffracted radiations (with an intensity above a threshold) from the moving grating 
over a range of angles and frequencies
$(\phi_p, \nu) \in [\phi_A,\phi_A + \Delta \phi_p] \times [\nu_A', \nu_A' + \Delta \nu']$.
We assume that the photodetector is operated at a distance equal to the visual range of a perfect reflector
and so it can only detect a reflected light from the grating if the intensity is 
higher or equal to that of a perfect reflector.
For an angle $\phi_p \in [\phi_A,\phi_A + \Delta \phi_p]$, the spectrally compressed radiation appears over a band 
$ [\nu' \in  [\nu'_{\rm min} (\phi_p), \nu'_{\rm min}(\phi_p) + \mathcal{W}']$ and the detector can collect the radiated energy from this band
if it is a part of $[\nu_A', \nu_A' + \Delta \nu']$.
This is illustrated in Fig.~\ref{Fig:detector:range}, where the red continuous curve (for $\nu'_{\rm min} (\phi_p)$)
and the blue dashed curve (for $\nu'_{\rm min}(\phi_p) + \mathcal{W}'$)
represent respectively the lower and upper edges of the region where the detector can receive a signal from the moving grating.

\medskip
The actual detection will also depend on the properties of the detector.
For instance, since the grating is rotating, the detector will be exposed to the bright light for only a finite time duration
and a detection may not happen if the duration is too short for the detector.
The exposure duration also depends on the angular velocity of the grating and the distance between the grating and the detector.
We can assume that a duration of a few milliseconds can be enough as the duration specifications of common photographic flashes are typically in the order of milliseconds.
The \emph{Morpho} butterflies flap their wings slowly and this should allow sufficient exposure time within a few hundred of meters.

%
%

\subsection{Cases where the angle of diffraction and the angle of observation are different}

\label{subsection:different:phi}

So far, we have assumed that the angle of diffraction is equal to the observation angle
(see Eq.~(\ref{eq1:observation:angle})),
although the spreading wavefront, from the field diffracted by a grating of finite size, 
means that the light diffracted into an order $p$ can also be seen at an observation angle $\phi_{\rm obs}$ 
which is not necessarily equal to the angle of diffraction $\phi_p$.
But in general we expect the brightest light to occur when the diffracted light is pointing directly toward the observer,
and so it can be reasonable to assume, in practice, that $\phi_p$ is equal $\phi_{\rm obs}$ in such situations.
However a notable exception to this rule can be the case of a grazing angle of diffraction.
Indeed, at a grazing angle of diffraction $\phi_p$, the angular spread has an asymmetric profile and even  
the peak intensity (for observation at a fixed distance from the grating) can occur at a viewing angle $\phi_{\rm obs}$  different from $\phi_p$.
For instance in Fig.~\ref{Fig:Fraunhofer:diffraction}, the peak intensity for the grazing angle of diffraction 
$\phi_p = 89.9^{\circ}$ appears at an angle $\phi = 87.54^{\circ}$ and it has deviated from the angle of diffraction $\phi_p$
 by $2.36^{\circ}$.

\medskip
When  $\phi_p$ is different from $\phi_{\rm obs}$,
the derivation leading to Eq.~(\ref{total:Doppler:shift}) can be modified as follows:
At reflection, the point $S$ can be seen as a light source moving at the speed $\bm{V}_p = - x_S \, \Omega_p \, \bm{z}$
and the velocity component in the observer's direction is given by $V_p \, \cos \phi_{\rm obs}$,
so that 
the total frequency shift can still be written in the form Eq.~(\ref{dispersion:eq1}) if the definition Eq.~(\ref{param:a})
of the parameter $A$ is changed to:
\begin{eqnarray}
\label{param:a:scale}
A  & = & - \frac{x_S \, \Omega_0}{2 \, c} \,
\left( 1+\frac{\cos \phi_{\rm obs}}{\cos \phi_p} \right).
\end{eqnarray}
Since the denominator term $\cos \phi_p$ in Eq.~(\ref{param:a:scale}) can be very small  at a grazing angle of diffraction $\phi_p$,
the absolute value of the parameter $A$ in Eq.~(\ref{param:a:scale}) can be much higher when 
$ \phi_p$ is different from $\phi_{\rm obs}$ (with $| \phi_p| > |\phi_{\rm obs}|$) than when 
$ \phi_p$ is equal to $\phi_{\rm obs}$. This can result in a stronger Doppler shift.
We also note that although Eq.~(\ref{angular:velocity}) shows that the angular velocity $\Omega_p$ of a diffraction order
at a grazing angle of diffraction $\phi_p$ can be much higher than $\Omega_0$,
the linear velocity $\bm{V}_p = - x_S \, \Omega_p \, \bm{z}$ has a vertical direction,
which is almost perpendicular to the direction of view of an observer when $ \phi_p = \phi_{\rm obs}$,
resulting in a smaller contribution to the Doppler shift than when $\phi_{\rm obs}$ is allowed to deviate from $\phi_p$
(toward the normal direction).

\medskip
The results in this section can also be relevant to multi-level micro/nanostructures such as butterfly wing scales~\cite{Mei:SM:2011}.
Each scale can be large enough to be modelled as a nanostructured diffraction grating.
The scales are in turn arranged as a microstructured array which covers a butterfly wing.
In the far field, constructive interference between the scale scatterers 
can occur in other directions than the original direction of the diffraction orders of the individual scales, i.e.,
$ \phi_p$ can be different from $\phi_{\rm obs}$.
%

%
%

%
%

\begin{figure}[htbp]
 \centerline{
\includegraphics[width=7cm]{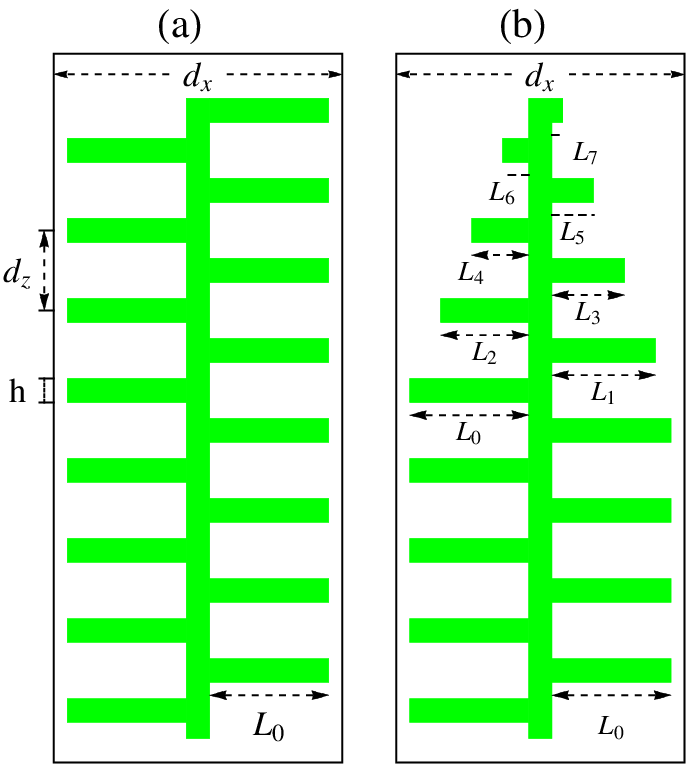}~
\includegraphics[width=0.9cm]{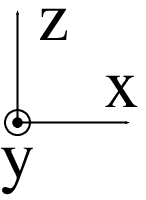}
}
\caption{The panels (a) and (b) show a unit cell of gratings with, respectively,  uniform alternating lamellae and apodized alternating lamellae.
}
\label{Fig:geometry:multilayer}
\end{figure}

%
%

%
%

\section{Numerical simulations}

\label{section:numerical:simulations}

%
%

%
For the numerical simulations, we use the two grating geometries shown in Fig.~\ref{Fig:geometry:multilayer}.
They are multilayer gratings consisting of alternating lamellae attached to a ridge structure.
The refractive index of the lamellae is $n = 1.56 + 0.06 \, i$ (chitin) and they are surrounded by air.
The grating periods in the $x$ and $z$ directions are $d_x = 746 \, {\rm nm}$ and $d_z = 207 \, {\rm nm}$.
The values of the lamella length $L_0$ and thickness $h$ are $L_0 = 308 \, {\rm nm}$ and $h = 62 \, {\rm nm}$.
The thickness of the ridge is $h_{\rm R} = 60 \, {\rm nm}$.
For the  grating in Fig.~\ref{Fig:geometry:multilayer}~(b), a linear apodization is applied to the length 
of the top 7 lamellae: $L_m = L_0 -  40 \, m$, for $m \in \{1, 2, \dots, 7 \}$.
The diffraction problem can be solved numerically using the finite element method presented in~\cite{Dossou:JCP:2006}
or a one-dimensional version of the finite element-based modal method described in~\cite{Dossou:JOSAA:2012}.

\medskip
The grating in Fig.~\ref{Fig:geometry:multilayer}~(b) models the nanostructures found in \emph{Morpho rhetenor} 
wing scales \cite[see Fig.~9]{Gralak:OE:2001}.
This grating can be seen as an apodized (or tapered) version of the one in Fig.~\ref{Fig:geometry:multilayer}(a).
Apodization is a technique where the strength of a grating is slowly reduced over a transition region, near an output or input end.
The profile of the \emph{Morpho rhetenor} grating has also some similarity with structures consisting of a regular photonic crystal whose 
interface is corrugated or indented. For instance with a photonic crystal consisting of lattice of rods, 
the corrugation can be introduced by removing some rods from the interface layer~\cite{Serebryannikov:PRB:2009}
or changing the cylinder radius in the interface layer~\cite{Maystre:OE:2001}.
Such structures, sometimes referred to as photonic crystal gratings, can combine the properties of photonic crystals 
(e.g., high reflectance or existence of surface modes inside a band-gap) and diffraction gratings (e.g., efficient scattering into higher diffraction orders)
to deliver some interesting diffraction properties.
We will compare the scattering properties of the two gratings in order to identify some possible advantages for the use of apodization 
in the nanostructures of \emph{Morpho} butterflies.

%
%

\begin{figure}[htbp]

\centerline{
\includegraphics[width=4.5cm]{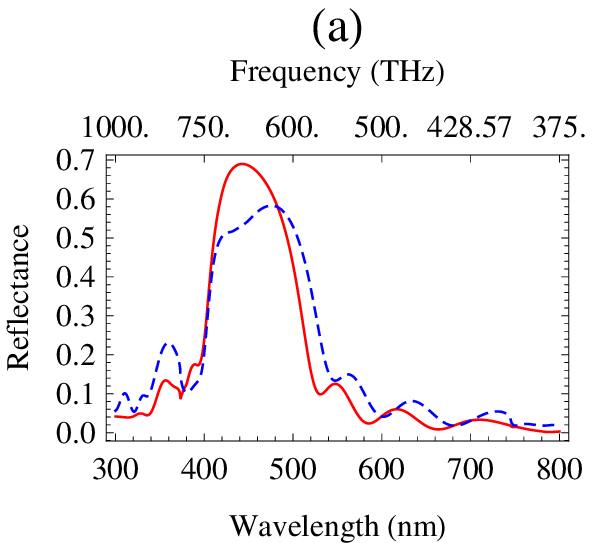}
}

\smallskip

\centerline{
\includegraphics[width=4.5cm]{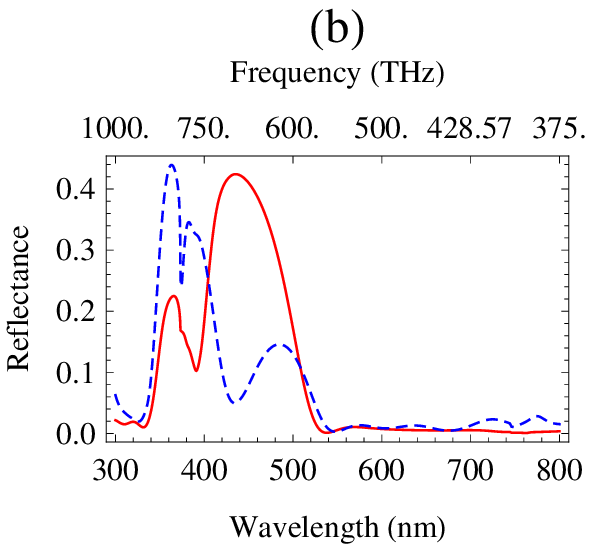}
}
\caption{The sub-figures (a) and (b) represent the total reflectance curves 
from the gratings shown respectively in Fig.~\ref{Fig:geometry:multilayer}~(a) and (b).
The red continuous curves and the blue dashed curves correspond respectively to incidence by $H_y$-polarized
and $E_y$-polarized plane waves.
The angle of incidence is normal.}
\label{reflectance:curves:1}
\end{figure}

%
%

\medskip
The total reflectance of the two gratings is shown in Fig.~\ref{reflectance:curves:1}, 
for normal incidence by $H_y$-polarized (red continuous curves) and $E_y$-polarized (blue dashed curves)
plane waves.
We can observe in Fig.~\ref{reflectance:curves:1}~(a) that the non-apodized grating has a high reflectance band
$\lambda \in [ 400 \, {\rm nm}, 510 \, {\rm nm} ]$.
The apodized grating can also reflect strongly the ultraviolet radiations in the wavelength range $\lambda \in [ 340 \, {\rm nm}, 400 \, {\rm nm} ]$.
We have noticed that the apodized grating reduces substantially the reflection into the specular order
(this can be beneficial to the reflectance into higher diffraction orders).
For instance the diffraction efficiency $e_0$
into the specular order is lower than 0.1\% for $\lambda \in [ 400 \, {\rm nm}, 500 \, {\rm nm} ]$ for both polarizations.
With the untapered grating, the diffraction efficiency $e_0$ can reach 5\% and 8\% respectively for the $H_y$ and $E_y$-polarizations,
over the same wavelength band.

\begin{figure}[htbp]

\centerline{
\includegraphics[width=7.5cm]{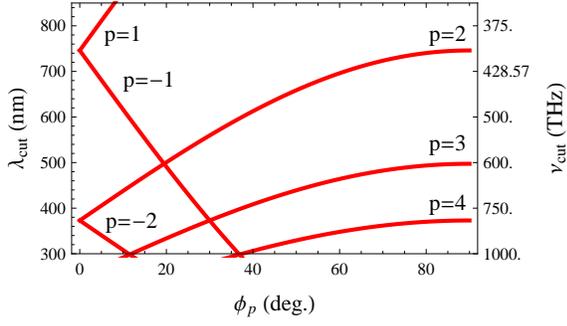}
}

\caption{The plots of the cut-off wavelength $\lambda_{\rm cut}$ (or cut-off frequency $\nu_{\rm cut}$) 
in Eqs.~(\ref{cut-off:p:positive}) and (\ref{cut-off:p:negative})
as a function of the angle of diffraction $\phi_p$,
for the diffraction orders $p=\pm1, \; \pm2, \; 3, \; 4$.
}
\label{cut:off:curves}
\end{figure}

\medskip
Figure~\ref{cut:off:curves} shows that the orders $p=-1,2,3$ can have a cut-off
in the wavelength range $\lambda \in [ 400 \, {\rm nm}, 510 \, {\rm nm} ]$
where the gratings can diffract strongly light from the visible spectrum.
Here we will carry out a numerical simulation for the order $p=3$, because its cut-off wavelength
at a grazing angle $\phi_{\rm obs}$ occurs within the band $\lambda \in [ 400 \, {\rm nm}, 510 \, {\rm nm} ]$.
For instance, when the angle of diffraction is set to $\phi_p = 89^{\circ}$, the cut-off wavelength
of the diffraction orders $p$=1, 2, 3 and 4 are respectively
$\lambda_{\rm cut}$ = 1491.89~nm, 745.943~{\rm nm}, 497.295~nm and 372.972~nm.
The order $p=4$ can be relevant for a study of the scattering in the ultraviolet band. 

\begin{figure}[htbp]

\centerline{
\includegraphics[width=8cm]{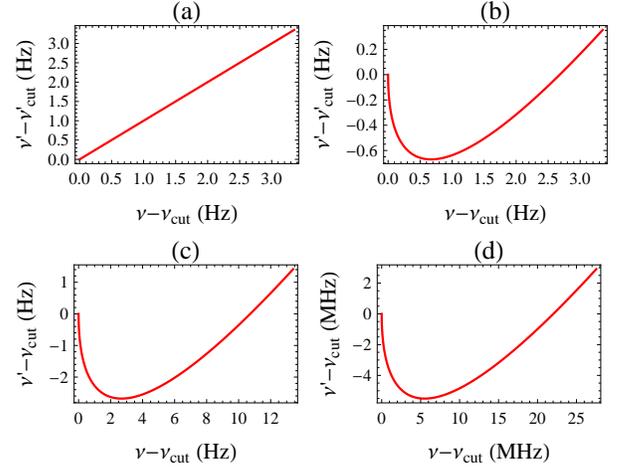}
}
\caption{
The panel (a) shows a Doppler dispersion curve
from a flat mirror under a translational motion (specular order).
The panels  (b), (c) and (d) show respectively the non-specular Doppler dispersion curves (order $p=3$)
from a grating under a translational motion,
 under a  rotational motion with $\phi_p = \phi_{\rm obs}$
and  under a  rotational motion with $\phi_p \neq \phi_{\rm obs}$.}
\label{Doppler:dispersion:curves}
\end{figure}

\medskip
For the numerical calculation of the Doppler shift  perceived by an observer watching a point $S$ on the grating interface, 
the linear velocity of $S$ is set to $V = -10 \, \rm{m/s}$, along the direction $\bm{z}$.
This velocity $V$ is supposed to be approximately in the same order of magnitude as the wing movement of a butterfly in flight.
%
%
%
The computed Doppler dispersion curves are shown in Fig.~\ref{Doppler:dispersion:curves}.
 Figure~\ref{Doppler:dispersion:curves}(a) shows the observed frequency from the specular order $p=0$ (or a moving flat reflector).
A specular order does not have a cut-off frequency and so the parameter $\nu_{\rm cut}$ can have an arbitrary value;
the choice of an arbitrary value for $\nu_{\rm cut}$ (or $\phi_0$) does not have any visually noticeable impact 
on the shape of the curve in  Fig.~\ref{Doppler:dispersion:curves}(a) since, from Eq.~(\ref{dispersion:eq1a:linear}), we have 
$\nu' - \nu'_{\rm cut} = (\nu - \nu_{\rm cut} ) \, (1 + 2 \, A \, \cos \phi_0) \approx  (\nu - \nu_{\rm cut} )$,
as $|A| = |V /c| = 3.333 \times 10^{-8} \ll 1$.
%

%
%

%
%

\medskip
 Figures~\ref{Doppler:dispersion:curves}(b) and (c) show the perceived frequency,
from the diffracted order $p = 3$,
respectively,  under a translational motion (see Eq.~(\ref{dispersion:eq1a:linear}))
and under a rotational motion with $\phi_3 = \phi_{\rm obs}$.
Both curves exhibit a flat band where the Doppler effect can be expected to amplify the perceived illumination.
Note that in the case of a translational movement, the frequency shift at the cut-off frequency $\nu_{\rm cut}$ is not zero,
accordingly the shifted frequency $\nu'_{\rm cut} = \nu_{\rm cut} \, ( 1 + V \, \cos \phi_3 / c)$
is used in the label $(\nu' - \nu'_{\rm cut})$ for the vertical axis in  Fig.~\ref{Doppler:dispersion:curves}(b) 
(a similar rule is also applied to Fig.~\ref{Doppler:dispersion:curves}(a)).
The value of the cut-off frequency $\nu_{\rm cut}$ depends on the diffraction angle $\phi_p$ 
(see  Eqs.~(\ref{cut-off:p:positive})-(\ref{cut-off:p:negative}) and Fig.~\ref{Fig:cutoff:order:p}),
but the asymptotic approximations Eqs.~(\ref{dispersion:approx1:linear}) and (\ref{dispersion:approx1})
shows that, near the cut-off, the shape of the Doppler dispersion curves in  Fig.~\ref{Doppler:dispersion:curves}(b) and (c)
is almost independent from $\phi_p$.
The non-linear effect from the Doppler shift is stronger in  Fig.~\ref{Doppler:dispersion:curves}(c) 
than in Fig.~\ref{Doppler:dispersion:curves}(b), as the  minimum frequency occurs further away from the cut-off.
Indeed, with  Fig.~\ref{Doppler:dispersion:curves}(c) we have $\nu_{\rm min} - \nu_{\rm cut} \approx 2 \, A^2 \, B_3 = 2.681\, \rm{Hz}$ (see Eq.~(\ref{nu:min}))
and we can verify that we get
$\nu_{\rm min} - \nu_{\rm cut} \approx A^2 \, B_3 /2 = 0.670\, \rm{Hz}$
with Fig.~\ref{Doppler:dispersion:curves}(b).
The frequency separation $(\nu_{\rm min} - \nu_{\rm cut})$ is four times larger in  Fig.~\ref{Doppler:dispersion:curves}(c) 
than in  Fig.~\ref{Doppler:dispersion:curves}(b) because the square root term in the asymptotic approximation
Eq.~(\ref{dispersion:approx1}) for a rotational motion
is multiplied by $2 \,  A$ while it is only multiplied by $A$ in the asymptotic relation Eq.~(\ref{dispersion:approx1:linear})
for a translational motion.
%

%
%

%
%

\medskip
The perceived diffracted light at the frequency $\nu'_{\rm min}$ is spectrally compressed.
In order to compare the visual range of a moving grating to that of a perfect flat reflector,
we have used a heuristic approach to model the propagation of spectrally compressed waves.
For a given diffraction angle $\phi_p$,
the field associated with a spectrally compressed band 
$\nu' \in  [\nu'_{\rm min}, \nu'_{\rm min} + \mathcal{W}']$
is approximated by a plane wave with a frequency $\nu'_{\rm min}$,
an angle of propagation equal to $\phi_p$
and a power flow $\mathcal{P}_{A}$ equal to the frequency-averaged power over the observed frequency band 
$\nu' \in  [\nu'_{\rm min}, \nu'_{\rm min} + \mathcal{W}']$
(the total power over $[\nu'_{\rm min}, \nu'_{\rm min} + \mathcal{W}']$ is given by  Eq.~(\ref{total:energy:p:prime})).
Since the maximum distance, at which the light reflected by a perfect flat mirror can be seen, occurs at a normal incidence,
we assume that another plane wave with a power $\mathcal{P}_{B}$ is propagating away from a perfect reflector at normal incidence.
By following the example of Fig.~\ref{Fig:Fraunhofer:diffraction},
we can use the  Fraunhofer diffraction formula  Eq.~(\ref{Fraunhofer:formula}) to find out how large the ratio 
$\mathcal{P}_{A}/\mathcal{P}_{B}$ needs to be so that 
a plane wave with a power $\mathcal{P}_{A}$, which propagates at an angle $\phi_p$, can be seen 
at a greater distance than a plane wave with a power $\mathcal{P}_{B}$, which propagates in the normal direction of the grating.
Once the required value of the ratio $\mathcal{P}_{A}/\mathcal{P}_{B}$ is determined, we can find 
the corresponding bandwidth $\mathcal{W}'$ of the compressed light by solving the power ratio relation Eq.~(\ref{power:ratio}),
where $\mathcal{D}_p'$ is set to $\mathcal{D}_p' = \mathcal{P}_{A}/\mathcal{P}_{B}$.
If the bandwidth $\mathcal{W}'$ is large enough (to be physically meaningful)
then we can expect the moving grating to have a comparable (or even a larger) 
visual range than a perfect reflector.

\medskip
For an application of the heuristic approach, we consider the case of Fig.~\ref{Doppler:dispersion:curves}(c).
We can verify that if a monochromatic plane wave (at a frequency $\nu'_{\rm min}$)
emerges from a finite grating of length $L = 200 \, \mu{\rm m}$
at an angle $\phi_3 = 60^{\circ}$, for instance, it can be visible beyond the visual range of a perfect flat reflector 
if $ \mathcal{P}_{A}$ is higher or equal to $2.5 \, \mathcal{P}_{B}$.
We note that if a plane wave has a power flow $\mathcal{P}_{B}$ at a normal propagation, for a propagation at $\phi_3 = 60^{\circ}$,
its power flow across the grating becomes  $\mathcal{P}_{B} \, \cos \phi_3 = 0.5 \, \mathcal{P}_{B}$,
which is 5 times smaller than $ \mathcal{P}_{A} = 2.5 \, \mathcal{P}_{B}$.
Here the values of the variables $\mathcal{D}_p'$, $|\rho_{p}|$, $\phi_p$
and $(\nu_{\rm min} - \nu_{\rm cut})$ in Eq.~(\ref{power:ratio}) are
$\mathcal{D}_p' = 2.5$,
$|\rho_{p}| = |\rho_3| = 7.0 \times 10^{-8}$,
$\phi_p = \phi_3 = 60^{\circ}$
and
$(\nu_{\rm min} - \nu_{\rm cut}) = 2.681 \, {\rm Hz}$,
so that we obtain
$\mathcal{W}' = 16 \, (\nu_{\rm min} - \nu_{\rm cut}) \,
|\rho_{p}^4| \, \cos^2 \phi_p / \mathcal{D}_p'^2
= 4.1397 \times 10^{-29} \, {\rm Hz}$.
With the present example, the bandwidth  $\mathcal{W}'$ is very small compared to the frequency 
$\nu'_{\rm min}$ and its physical significance is uncertain.

\medskip
The results above suggest that the Doppler shift from the point $S$ on the grating is too small to produce a significant spectral compression
when the point $S$ has a linear speed of $V = -10 \, \rm{m/s}$ and $\phi_p = \phi_{\rm obs}$.
However, for a rotating grating, although the point $S$ is moving physically at a speed 
$V = - x_S \, \Omega_0= -10 \, \rm{m/s}$ (see Eq.~(\ref{speed:V:0})),
we note that, with a diffracted field of order $p \neq 0$, the point $S$ simulates a source point moving at a speed
$V_p = V \, \cos \phi_0 / \cos \phi_p$ (see Eqs.~(\ref{angular:velocity}) and (\ref{speed:V:p}))
and so the absolute value of $V_p$ can be substantially higher than $10 \, \rm{m/s}$
if $\cos \phi_0 \gg \cos \phi_p$, i.e., if $\phi_p$ is close enough to $\pm \pi/2$.
We also note that, for the calculation of the Doppler shift, it is the velocity component in the observer's direction, i.e., $V_p \, \cos \phi_{\rm obs}$,
that is relevant.
When $\phi_{\rm obs}$ is equal to $\phi_p$, the denominator term $\cos \phi_p$ in the expression for $V_p$ is cancelled 
and the Doppler shift will not be amplified when $\phi_p$ is close to $\pm \pi/2$.
But the Doppler shift can be very strong at a grazing angle of diffraction $\phi_p$
when the grating is in a  rotational motion with $\phi_p \neq \phi_{\rm obs}$.

\medskip
Indeed with a grazing angle of diffraction $\phi_p$, as pointed out in Section~\ref{section:rotational:motion}\ref{subsection:different:phi}, 
it can be reasonable to consider situations where 
the angle of diffraction $\phi_p$ is different from the angle of observation $\phi_{\rm obs}$.
The interference between neighboring scales of a butterfly wing can further increase this angular deviation
(for example most of the non-evanescent diffraction orders of the scale lattice will propagate
in directions which deviate away from the grazing direction).
However we do not have a simple formula for evaluating such a deviation,
and for an illustration purpose, the value of the angular deviation is set arbitrarily to $30^{\circ}$ in this work
(the main observations will still hold for other non-zero deviation values).
The Doppler dispersion curve in Fig.~\ref{Doppler:dispersion:curves}(d) is obtained by
using $\phi_3 = 89.99^{\circ}$  and $\phi_{\rm obs} = \phi_3 - 30^{\circ} = 59.99^{\circ}$.
We can see that, in Fig.~\ref{Doppler:dispersion:curves}(d), the Doppler frequency shift is stronger near the cut-off frequency 
than in Figs.~\ref{Doppler:dispersion:curves}(b) and (c)
(in Fig.~\ref{Doppler:dispersion:curves}(d) the frequency is in megahertz unit MHz).
This is due to the fact that the parameter $A$ for Fig.~\ref{Doppler:dispersion:curves}(d) is given by Eq.~(\ref{param:a:scale}),
i.e., $A = -4.777 \times 10^{-5}$, and it is about 1433 times larger than the value of the parameter $A$ 
(from Eqs.~(\ref{param:a:linear}) and (\ref{param:a}))
used in  Figs.~\ref{Doppler:dispersion:curves}(b) and (c), i.e., $A = - 3.333 \times 10^{-8} $.
%

%
%

\begin{table}
\begin{tabular}{lllll}
\multicolumn{1}{c}{$\phi_3$} & 
$\nu_{\rm min}$ & 
\multicolumn{1}{c}{$|V_3|$} & 
\multicolumn{1}{c}{$|\rho_3|$} & 
\multicolumn{1}{c}{$\mathcal{W}' $} \\
(deg.) & 
(\rm{THz}) & 
\multicolumn{1}{c}{(m/s)} & 
 & 
\multicolumn{1}{c}{(Hz)} \\ \hline
$89.0$ & \hspace*{-0.4cm} 603.263 & \hspace*{-0.4cm} $0.00116$ & \hspace*{-0.4cm} $7.68 \times 10^{-6}$ & \hspace*{-0.3cm} $4.71 \times 10^{-23}$ \\
$89.9$ & \hspace*{-0.4cm} 603.218 & \hspace*{-0.4cm} $0.1101$ & \hspace*{-0.4cm} $7.88 \times 10^{-5}$ & \hspace*{-0.3cm} $2.62 \times 10^{-17}$ \\
$89.99$ & \hspace*{-0.4cm} 603.217 & \hspace*{-0.4cm} 10.95 & \hspace*{-0.4cm} $7.90 \times 10^{-4}$ & \hspace*{-0.3cm} $2.61 \times 10^{-11}$ \\
$89.999$ & \hspace*{-0.4cm} 603.218 & \hspace*{-0.4cm} 1094. & \hspace*{-0.4cm} $7.82 \times 10^{-3}$ & \hspace*{-0.3cm} $2.51 \times 10^{-5}$ \\
$89.9999 $ & \hspace*{-0.4cm} 603.272 & \hspace*{-0.4cm} $109418.$ & \hspace*{-0.4cm} $7.15 \times 10^{-2}$ & \hspace*{-0.3cm} $17.55$
 \\ \hline
\end{tabular}
\caption{Values of $\nu_{\rm min}$, $|V_3|$, $|\rho_3|$ and $\mathcal{W}' $
as $\phi_3$ approaches the right angle.
The diffraction coefficient  $\rho_3$ at $\nu_{\rm min}$ corresponds to an incidence by $H_y$-polarized 
plane waves over the apodized grating.}
\label{table:param:W}
\end{table}

%
%

\medskip
The Table~\ref{table:param:W} shows the frequency $\nu_{\rm min}$, the linear velocity $V_3 = V \, \cos \phi_0 / \cos \phi_3$,
the diffraction coefficient $|\rho_3|$ at $\nu_{\rm min}$ 
and the computed  frequency bandwidth $\mathcal{W}'$ 
(where the visual range of the moving grating may be equal or higher than that of a perfect reflector)
when the angle of diffraction $\phi_3$ is set to:
$\phi_3^{(1)} = 89.0^{\circ}$, 
$\phi_3^{(2)} = 89.9^{\circ}$,
$\phi_3^{(3)} = 89.99^{\circ}$, 
$\phi_3^{(4)} = 89.999^{\circ}$
and
$\phi_3^{(5)} = 89.9999^{\circ}$
(with $\phi_{\rm obs} = \phi_3^{(i)} - 30^{\circ}$).
We can see that the linear speed $V_3$ associated with the rotating diffracted field can take very large values 
when $\phi_3$ approaches a right angle.
In order to determine $\mathcal{W}'$, we have first used the Fraunhofer formula  Eq.~(\ref{Fraunhofer:formula}) to find
${\mathcal{D}_p'}^{(i)} = \mathcal{P}_{A}/\mathcal{P}_{B}$  such that a plane wave emerging at an angle $\phi_3 = \phi_3^{(i)}$ 
from the grating, with a radiated power $\mathcal{P}_{A}$,
can be seen at a greater distance than a plane wave with a radiated power $\mathcal{P}_{A}$
which propagates away from the grating in a normal direction; 
for $i=1,\dots,5$ we have:
${\mathcal{D}_p'}^{(1)} = 15$, 
${\mathcal{D}_p'}^{(2)} = 2$, 
${\mathcal{D}_p'}^{(3)} = 0.2$, 
${\mathcal{D}_p'}^{(4)} = 0.02$
and
${\mathcal{D}_p'}^{(5)} = 0.002$.
We can then obtained $\mathcal{W}'$ from Eq.~(\ref{power:ratio}) by setting the value of $\mathcal{D}_p'$
to ${\mathcal{D}_p'}^{(i)}$.
The values of $\mathcal{W}'$ in Table~\ref{table:param:W} increase rapidly as $\phi_3$ approaches a right angle
and so we can expect that, at high grazing angles of diffraction, the moving grating can 
reflect a bright light which can be seen at a greater distance than the light reflected by a perfect reflector.

\medskip
Since a rotating grating can produce a stronger spectral compression
at a grazing  angle of diffraction  $\phi_p$ (with $\phi_p \neq \phi_{\rm obs}$),
it can be desirable to use a diffraction grating which can deliver a relatively high 
diffraction efficiency $e_p$ (or diffraction coefficient $\rho_p$)
when $\phi_p$ is close to a right angle.
The results in Table~\ref{table:param:W} correspond to an incidence over the apodized grating 
by $H_y$-polarized plane waves.
When we repeat the same calculations with the non-apodized grating,
the values of the calculated diffraction coefficients $|\rho_3|$ were about 50\% 
(more precisely between 43\% and 48\%) of the values obtained with the apodized grating.
For incidence by $E_y$-polarized plane waves, the performance of the non-apodized grating
at grazing  angles of diffraction is even worse since the values of the diffraction coefficients $|\rho_3|$
are between 3.7\% and 8.9\% of the value given by the apodized grating.
The superior behavior of the tapered grating shows
that it is better-suited for the purpose of spectral compression at grazing angles of diffraction.
%

%
%

\section{Conclusion}

A light ray emerging from a point $S$ on a grating, illuminated by a polychromatic diffuse light, 
can be seen as a superposition of the diffracted fields from a multitude of incident rays which are converging toward the point $S$.
For a given non-specular diffraction order $p$, the angle of incidence of each incident ray is frequency-dependent
and so, as the grating moves, each incident ray (and its diffracted ray) can experience a different Doppler effect.
The most dramatic Doppler effect occurs near a Wood anomaly, where the Doppler dispersion curve from a receding point $S$ 
can exhibit a flat band.
The incident frequencies inside a flat band can be Doppler-shifted into a narrower band of perceived frequencies
(spectral compression). From the principle of energy conservation,
the perceived light from the compressed frequency band can be brighter than the light diffracted by a stationary grating
or even brighter than the light reflected by a perfect passive reflector with a reflectance of 100\% 
(since the spectral compression factor can be arbitrarily large).

\medskip
In particular, this means that a moving grating, under the sunlight, can be seen as a kind of a \emph{super reflector}.
Our work also leads to the question as to whether the intense blue flash, which can be perceived when the \emph{Morpho rhetenor} butterfly is flying under the sunlight,
can be attributed to a spectral compression mechanism.
Although this can be possible according to our theoretical results,
a proper answer to this question can require further research (e.g., a direct observation) and can be the subject of a future work.
Finally, although we have considered the simplified model of a one-dimensional grating, with the incident and diffracted rays lying in a plane perpendicular to the grating grooves,
our results can still hold for more general grating geometries, as they are mainly based on the strong angular dispersion near a Wood anomaly.

%
%

%

%
%

\section*{Acknowledgments}
This research was supported by the Australian Research Council (ARC) 
Centre of Excellence for Ultrahigh Bandwidth Devices for Optical Systems 
(project number~CE110001018).
It is a great pleasure to thank Professor Lindsay C. Botten
and Associate Professor Christopher G. Poulton for many interesting discussions.
We are also grateful to Dr Paul K. Agboati for his inspiring comments.

%
%

%
%

%
\end{document}